\documentstyle[12pt,pic98pro,epsf,epsfig,wrapfig]{article}
\begin{document}
\title{ 
PHOTON STRUCTURE AND GAMMA-GAMMA PHYSICS
}
\author{
D.J.Miller,                             \\
{\em Department of Physics and Astronomy, University College
London,}\\
{\em Gower Street, London WC1E 6BT, UK}
}
\maketitle
\baselineskip=14.5pt
\begin{abstract}
The LEP experiments are making real progress in understanding the
structure of the photon, though the results do not yet give such clear
demonstrations of QCD in action as the proton structure has done.  Other
new results are reported, including QED related effects and 
$\gamma \gamma \rightarrow Resonances$, from LEP and from CLEO II.
\end{abstract}
\baselineskip=17pt
\section{\bf Introduction}
\begin{wrapfigure}[6]{r}{9.0cm}
 \vspace{-0.2cm}
 \epsfig{file=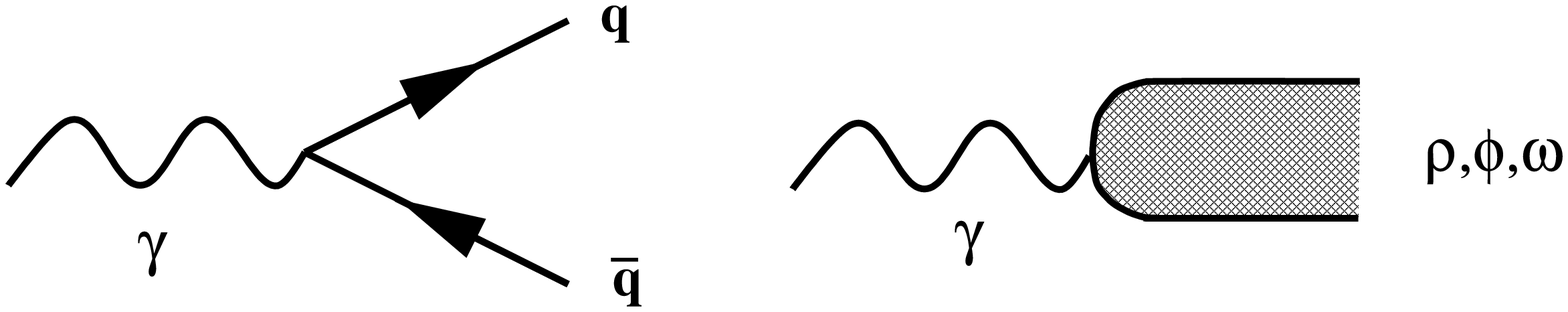,height=2.3cm,width=9.cm}
 \vspace{-0.7cm}
 \caption{The dual nature of the photon.}
 \label{fig:graph}
 \end{wrapfigure}

The photon is not a hadron -- it has fundamental direct couplings to all 
charged particles - {\em except} when it has already turned into a hadron 
before it interacts, see Figure~\ref{fig:graph}.  This duality gives rise to a wide range 
of phenomena that test both the perturbative and non-perturbative parts of QCD.
  Current experiments at LEP and CLEO II at Cornell are improving our 
understanding of photon structure and the properties of gamma-gamma resonances.

\begin{wrapfigure}[13]{r}{6.8cm}
 \vspace{-0.1cm}
 \epsfig{file=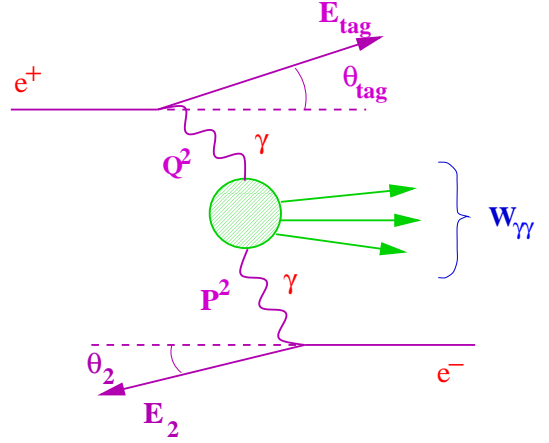,height=5.8cm,width=7.0cm}
 \vspace{-0.6cm}
 \caption{Singly tagged $\gamma\gamma$ event.}
 \label{fig:tag}
 \end{wrapfigure}

Figure~\ref{fig:tag} shows the generic Feynman graph for all $\gamma \gamma$ processes at 
an $e^+ e^-$ collider.  It is labelled to show one of the scattered beam 
leptons as a ``tagged'' and measured final state particle.  The results which 
follow come from untagged events 
as well as this singly tagged case.  If a scattered lepton is tagged then its 
four momentum transfer is usually well measured, and we define 
$Q^2=-q^2=2E_{beam}E_{tag}(1-\cos\theta_{tag})$.
If a scattered lepton is untagged then we know that its scattering angle was 
less than a few tens of milliradians and that $P^2\simeq 0$.

The invariant mass $W_{\gamma \gamma}$ can sometimes 
be well measured -- if all the 
final state particles are caught by the detector.  But in multihadron final 
states some of the particles are often missed, which means that both 
$W_{\gamma \gamma}$ and the Bjorken scaling variable 
$$x={Q^2\over Q^2+P^2+W^2_{\gamma\gamma}}\simeq 
    {Q^2\over Q^2+W^2_{\gamma\gamma}}$$
will be badly measured, and biased.  There is no alternative way of 
determining $W_{\gamma \gamma}$ from the initial state kinematics, as can 
be done for $W_{\gamma p}$ in electron-proton scattering at HERA, because 
both virtual $\gamma$s are drawn from broad bremsstrahlung-like spectra.  

Virtual photon beams bring two other disadvantages:  \\
i) the bulk of the $\gamma \gamma$ cross section is at very low values of 
$W_{\gamma \gamma}$~\cite{Kolan} so the effective luminosity for high energy 
processes is much less than at HERA.\\
ii) the longitudinal momenta of the two colliding photons are usually unequal,
giving a strong boost along the beam direction to the final state and 
accentuating the loss of particles from the well-measured region.

The prospect of eventually producing ``narrow band'' beams of real photons by 
Compton backscattering at a future linear collider has been extensively 
discussed elsewhere~\cite{Compton}.  Such a facility would raise the effective 
luminosity for interesting hard processes to at least the level of the 
$\gamma p$ luminosity at HERA, with a significant part of the spectrum within 
10\% of the peak value of $\sqrt{s_{\gamma \gamma}}$.  As well as allowing 
much of the physics discussed in the rest of this review to be done with 
better precision, a high energy ``Compton Collider'' could give important 
access to properties of Higgs and SUSY particles which could not be seen in 
other ways and would extend the range of constraints on the 
triple-gauge-boson couplings.  A low energy 
Compton Collider would be a unique facility 
for the study of meson resonances, certainly up to around the $\eta_c$, may 
even allow a chance to see the $\eta_b$~\cite{Borden}.

\section{\bf Photon Structure }

The differential cross section for singly tagged processes is~\cite{Kolan}
$${d^2\sigma(e\gamma\rightarrow ex)\over dxdQ^2} 
   ={2\pi\alpha^2\over xQ^4} 
  \left\{ \left(
             1+(1-y)^2
          \right)
          F^\gamma_2(x,Q^2)-y^2F_L^\gamma(x,Q^2)
   \right\}.
$$

This has been integrated over the azimuthal angle of
the unseen scattered 
lepton.  The second scaling variable 
$y=1-E_{tag}/E_{beam}(\cos^2\theta_{tag}/2)$
is much less than 1 if the threshold for the tagged electron is set at more 
than half the beam energy, as it normally must be at LEP to eliminate 
beam-associated background from the collider, so the $y^2 F_L$ contribution 
is expected to be unmeasurably small~\cite{Miller86}.  The main part of 
$e \gamma$ deep inelastic scattering is therefore driven by the structure 
function $F_{2}(x,Q^{2})$.
 
\subsection{The QED Structure of the Photon}

 \begin{wrapfigure}[17]{r}{9.cm}
 \vspace{-1.1cm}
 \epsfig{file=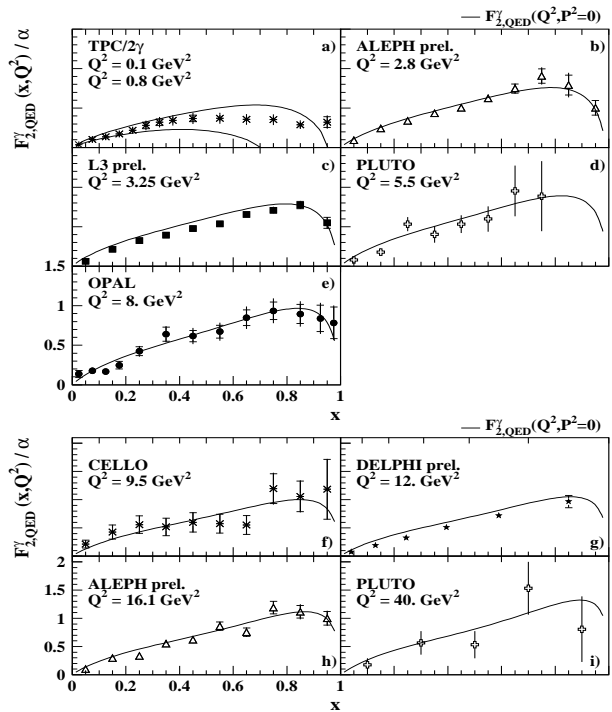,height=12.8cm,width=9.9cm,clip=}
 \vspace{-3.9cm}
 \caption{Measurements of $F_2^{QED}$.}
 \label{fig:f2qed}
 \end{wrapfigure}

All of the LEP experiments have confirmed the predicted shape and size of 
$F^{QED}_2$ in the singly-tagged $\gamma^* \gamma \rightarrow \mu^+ \mu^-$ 
process~\cite{F2mu}, Figure~\ref{fig:f2qed}.  
A recent L3 result~\cite{L3mu} even claims to see 
signs of the slight reduction in rate expected because the average value of 
$P^2$ is not exactly zero if the second tagged electron is only vetoed 
down to about 35 milliradians from the beam direction.  The universality of 
the tau coupling has been checked~\cite{tau}.  L3 has also 
measured the 
rates for untagged $\gamma \gamma \rightarrow e^+ e^-$, $\mu^+ \mu^-$ and 
$\tau^+ \tau^-$~\cite{L3QEDuntag}. 

\begin{wrapfigure}[10]{r}{5.6cm}
 \vspace{0.2cm}
 \epsfig{file=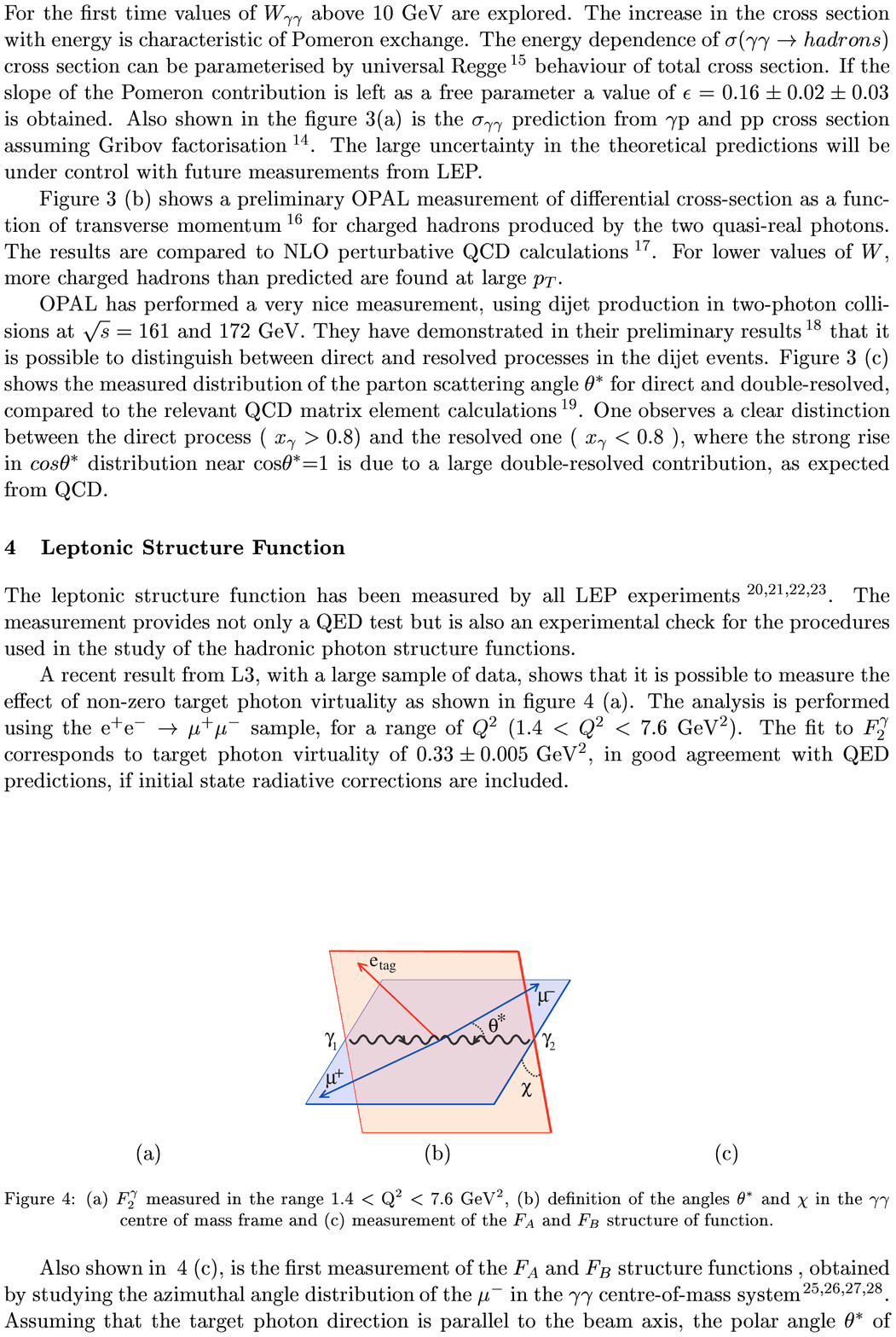,height=3.9cm,width=5cm,clip=}
 \vspace{-0.1cm}
 \caption{Definition of the angle $\chi$ in the
 $\gamma\gamma$ C. of M. frame.}
 \label{fig:tagplane}
 \end{wrapfigure}

An intriguing set of studies by three of the LEP experiments~\cite{FAFB} has 
also confirmed QED predictions for a pair of more subtle structure functions, 
$F^{QED}_A$ and $F^{QED}_B$, which relate to the azimuthal angle $\chi$ 
between the plane of the tagged electron and the plane of the two outgoing 
muons in the $\gamma \gamma$ C. of M., see Figure~\ref{fig:tagplane}.  
$F_A$ multiplies the 
$cos \chi$ dependence and comes from the interference between longitudinal 
and transverse photon scattering.  $F_B$ multiplies $cos 2\chi$ and is a 
transverse-transverse term.  

Apart from an apparent disagreement on 
sign conventions - an easy problem for the LEP inter-experiment 
$\gamma \gamma$ working group to sort out - the results agree with each other 
and with QED predictions. 

Measurement of $F^{QED}_A$ and $F^{QED}_B$ is interesting, not because it 
tests QED but because it proves that such measurements are possible at LEP.  
The challenge now is to do the same thing with hadronic final states, using 
the plane of pairs of jets to define the plane of the outgoing quarks instead 
of the dimuon plane.  QCD calculations of $F_A$ and $F_B$ involve many of the 
same operators as calculations of the unmeasurable $F_L$, which is expected 
to scale at lowest order in QCD.

\subsection{QCD Evolution of $F^{\gamma}_2 (x,Q^2)$ with $Q^2$ }

\begin{wrapfigure}[14]{r}{8.3cm}
 \vspace{-.7cm}
 \epsfig{file=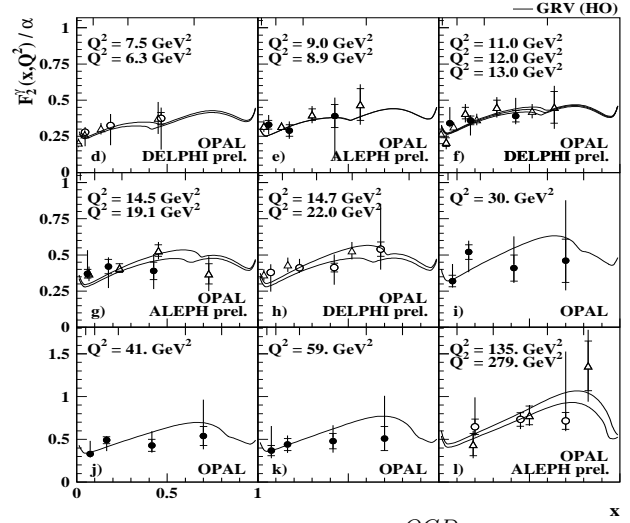,height=7.3cm,width=8.4cm,clip=}
 \vspace{-1.1cm}
 \caption{Measurements of $F_2^{QCD}$.  The curves are for the GRV 
parameterisations of the parton density functions~$^{\displaystyle {17)}}$.}
 \label{fig:f2qcd}
\end{wrapfigure}
This classic test of QCD~\cite{Witten} has been complicated by both 
theoretical and experimental problems.  The theoretical picture is 
complicated by the dual nature of the photon.  A purely perturbative 
treatment, starting from the direct $\gamma q$ coupling, gives singularities 
at low $x$.   
These can be dealt with by including the pre-existent hadronic 
part of the photon, but the predictive power of QCD is then severely 
undermined by the need for a nonperturbative description of the parton 
distribution in the initial object -- usually taken to be a vector meson 
with parton structure similar to that of the pion.  

\begin{wrapfigure}[15]{r}{8.3cm}
 \vspace{-.8cm}
 \epsfig{file=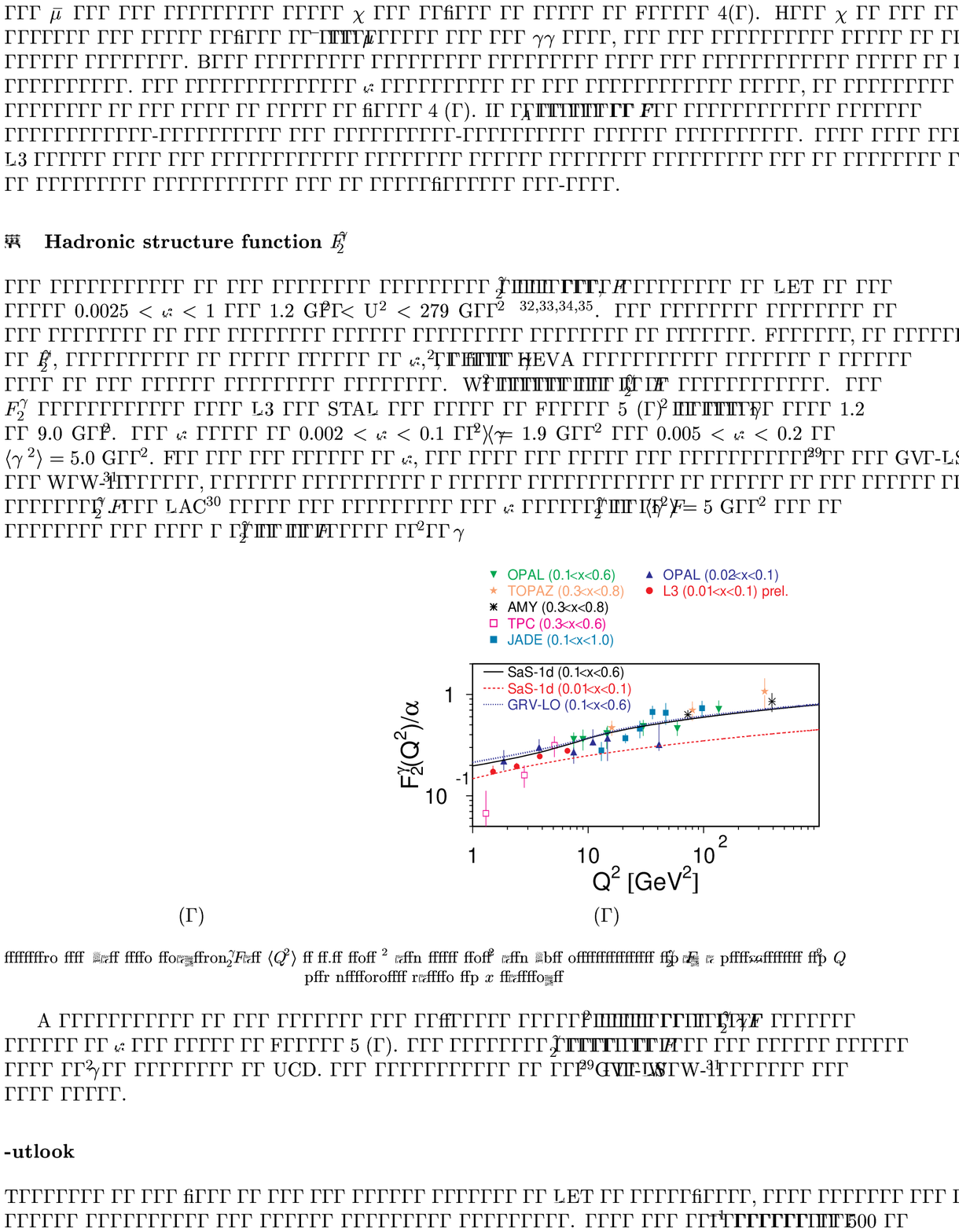,width=8.4cm,
 height=7.8cm,clip=} 
 \vspace{-0.9cm}
 \caption{$Q^2$ evolution of $F_2^{QCD}$.  The curves are for the SAS 
  parameterisations~$^{\displaystyle {17)}}$.}
 \label{fig:f2vsq2}
\end{wrapfigure}
\begin{figure}[t]
 \epsfig{file=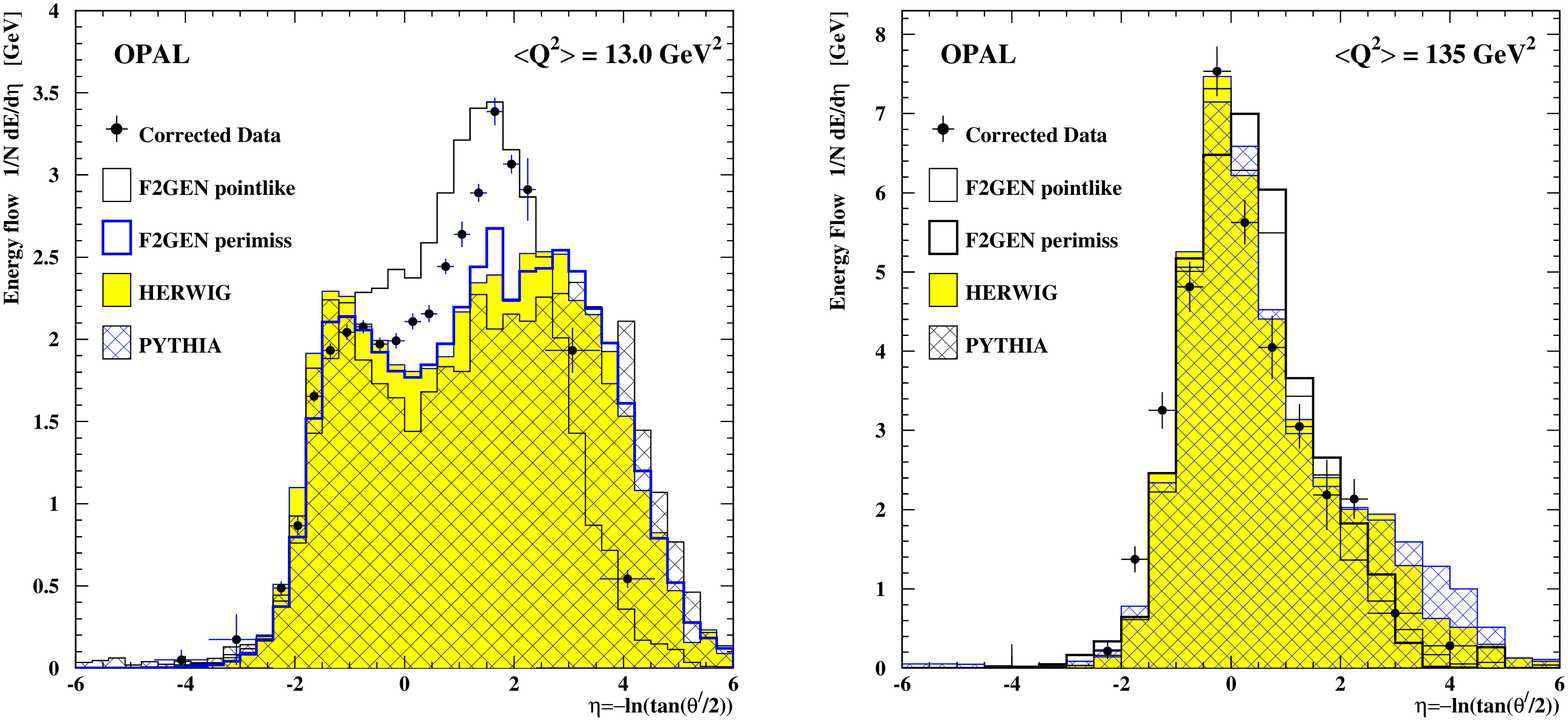,height=7.2cm,width=15.6cm}
   \vspace{-0.3cm}
  \caption{\label{fig:eflow}(a) Energy flow at $<Q^2>=$13.0 GeV$^2$ and
  (b) $<Q^2>=$135 GeV$^2$. 
}
\end{figure}
\begin{wrapfigure}[17]{r}{7.6cm}
 \vspace{-0.4cm}
 \epsfig{file=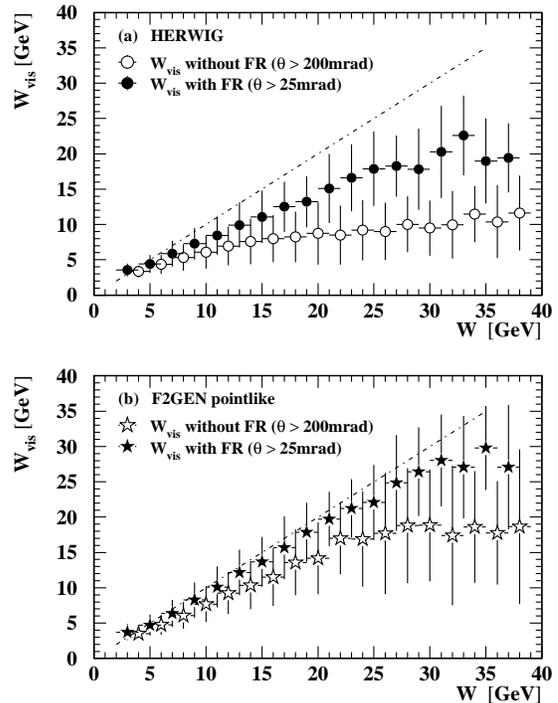,height=9.7cm,width=7.7cm,clip=}
 \vspace{-0.9cm}
 \caption{$W-W_{vis}$ correlation.}
 \label{fig:wwvis}
 \end{wrapfigure}

Nevertheless, it has 
been argued~\cite{Fraser,LEP200} that some sensitivity to the QCD evolution 
of the direct coupling does remain if measurements can be made over a wide 
range of $Q^2$. 
The LEP experiments~\cite{F2LEP} are making progress in 
measuring $F^{\gamma}_2 (x,Q^2)$ 
(Figure~\ref{fig:f2qcd}), and in seeing its expected 
nonscaling growth with $\ln Q^2$ 
(Figure~\ref{fig:f2vsq2}), but the error 
bars are enormous compared with those for $F^{p}_2$ from HERA, from SLAC or 
from muon beams.  This is partly statistical, due to the softness of the 
virtual photon spectrum at LEP, but the major part of the error is systematic,
 due to uncertainties in reconstructing the mass $W_{\gamma \gamma}$ of the 
hadronic final state.  The LEP detectors have good hadron tracking out to a 
pseudorapidity $|\eta| \simeq 2.3$, but beyond that they only have 
electromagnetic calorimetry which samples the hadronic final state but does 
not measure all of the hadronic energy.  It is therefore necessary to rely 
upon Monte Carlo simulation in an unfolding procedure~\cite{BlobelKd} 
which corrects for the distorted $x$ distribution 
caused by the biased measurement of $W_{\gamma \gamma}$.  Different Monte 
Carlo programs~\cite{MC} give very different forward energy flows 
(Figure~\ref{fig:eflow}a and b), so 
correlations (Figure~\ref{fig:wwvis}) between 
the observed $W_{vis}$ and the true value of 
$W$ depend upon both whether the Forward Detector hadronic energy is 
sampled at all (``without FD'', ``with FD'') and which 
Monte Carlo is used.  The experimental error bars in 
Figures~\ref{fig:f2qcd} and~\ref{fig:f2vsq2} are dictated primarily by 
the choice of the set of Monte Carlo programs used for input to the unfolding.

The all-LEP working group is making good progress on improving this 
situation.  Better Monte Carlo generators are being provided -- though there 
is always a danger here~\cite{MillerEg}.  Many generators assume a particular 
parameterisation of the parton density functions inside the photon.  If the 
parton-showering part of such a program is tuned to fit data like that in 
Figure~\ref{fig:eflow}a), then there is a chance 
that the unfolded $F^{\gamma}_2 (x,Q^2 )$ 
will not be independent from the assumed parameterisation.  Fortunately, new 
techniques are also being devised which will make the final unfolded result 
less sensitive to the particular Monte Carlo used. 

At large values of $Q^2$ the measured hadronic energy flow 
(Figure~\ref{fig:eflow}b)  is 
much better described because the hadrons recoil against the larger 
transverse momentum of the tagged electron, so a larger fraction of them is 
caught by the central trackers.  But at high $Q^2$ and high $x$ there is 
another problem -- the treatment of the charm mass, both in the Monte Carlo 
models and in the published parton density functions~\cite{pdfs}.  Some 
treat charm as massless but impose a sudden threshold cut, giving a totally 
unphysical sharp edge in  $F^{\gamma}_2 (x)$.  Some take 
$m_c =1.3 GeV/c^2 $, others have $m_c =1.7GeV/c^2$.  
At the highest $x$ values 
there may also be difficulties with simple factorisation of the effective 
$\gamma \gamma$ luminosity.  This must all be sorted out before we can 
deliver the final LEP measurement of the $Q^2$ evolution. 

LEP should generate three times more luminosity in 1998--2000 than has yet 
been analysed for photon structure by any single experiment.  It will be at 
higher energies -- giving an increased reach in $Q^2$ -- and the four LEP 
experiments are collaborating well in understanding the problems so that 
their eventual results may be combined safely.  Assuming we can beat down 
the systematic errors to match these statistics then a definitive measurement 
of the $Q^2$ evolution should be possible.  It may even be possible to 
extract a new value of $\Lambda_{\overline{MS}}$, though 
probably only within the 
framework of a particular parameterisation of $F^{\gamma}_2 (x,Q^2)$ like GRV 
or SaS.  To get a truly model independent measurement may require the Compton 
Collider~\cite{Millervogt}.

\subsection{Low {\em x} QCD Evolution of $F^{\gamma}_2 (x,Q^2)$ }

 \begin{wrapfigure}[17]{r}{8.2cm}
 \vspace{-0.8cm}
 \epsfig{file=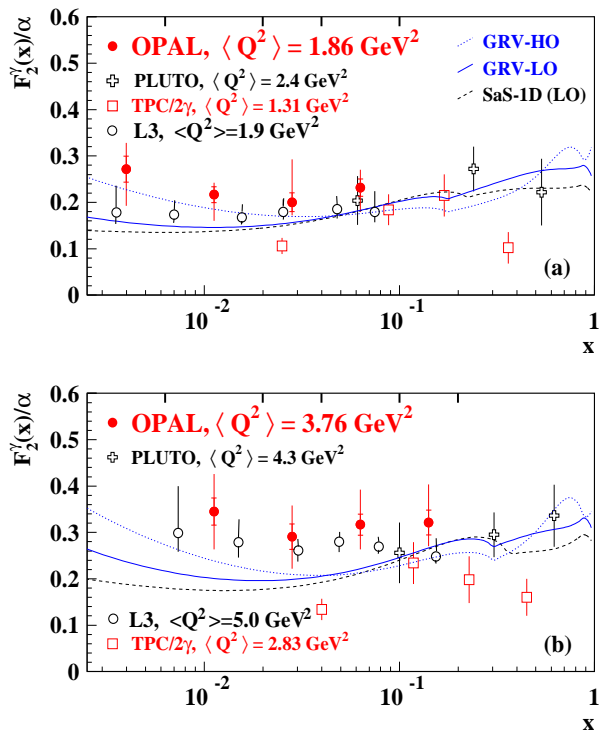,height=9.8cm,width=8.3cm,clip=}
 \vspace{-0.9cm}
 \caption{Measurements of $F_2^\gamma$ at low $x$.}
 \label{fig:lowx}
 \end{wrapfigure}

\begin{figure}[p]
 \epsfig{file=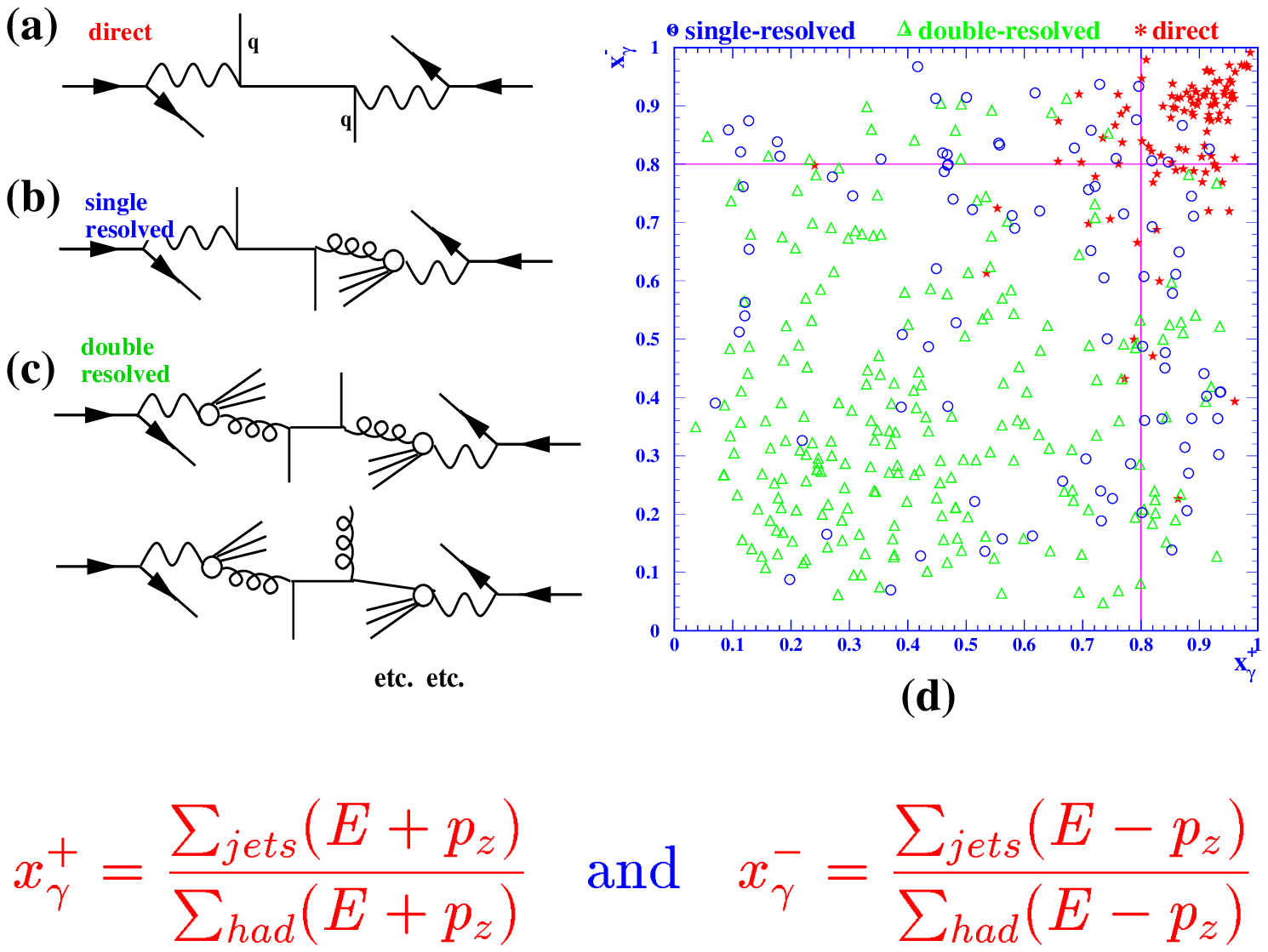,
         height=11.6cm,width=17.3cm,clip=}
   \vspace{-0.9cm}
  \caption{\label{fig:scat} Separation of the different components
of $\gamma \gamma$ scattering in the Monte Carlo sample from the 
PYTHIA program.~$^{\displaystyle {27)}}$}
 \vspace{+1.6cm}
 \hspace{-0.cm}
 \begin{minipage}[t]{0.45\linewidth}
   {\epsfig{file=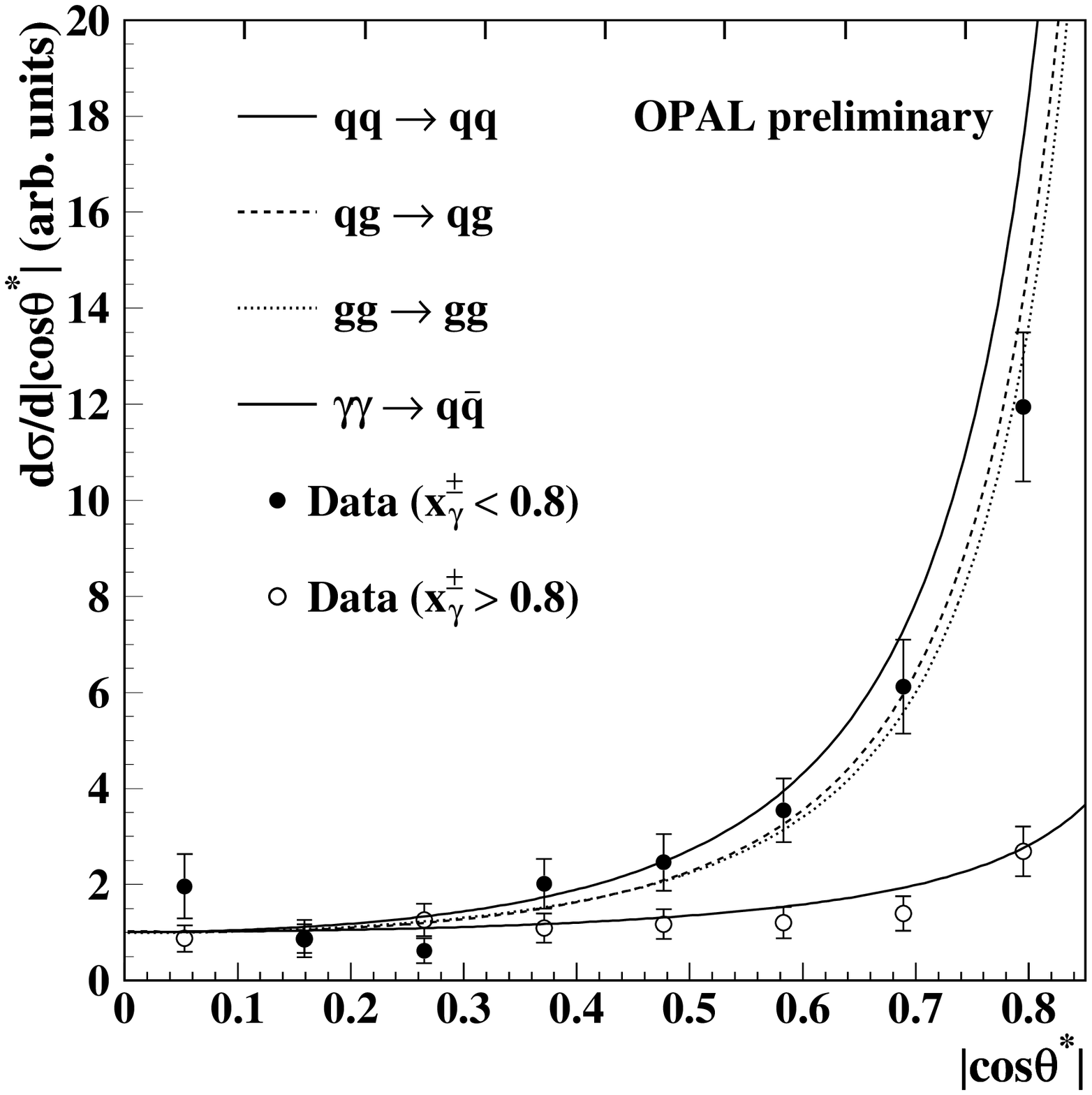,height=5.4cm,width=0.9\linewidth}}
  \caption{\label{fig:angular} Angular distribution in the dijet 
C. of M.~$^{\displaystyle {27)}}$}
 \end{minipage}\hfil
 \begin{minipage}[t]{0.48\linewidth}
   {\epsfig{file=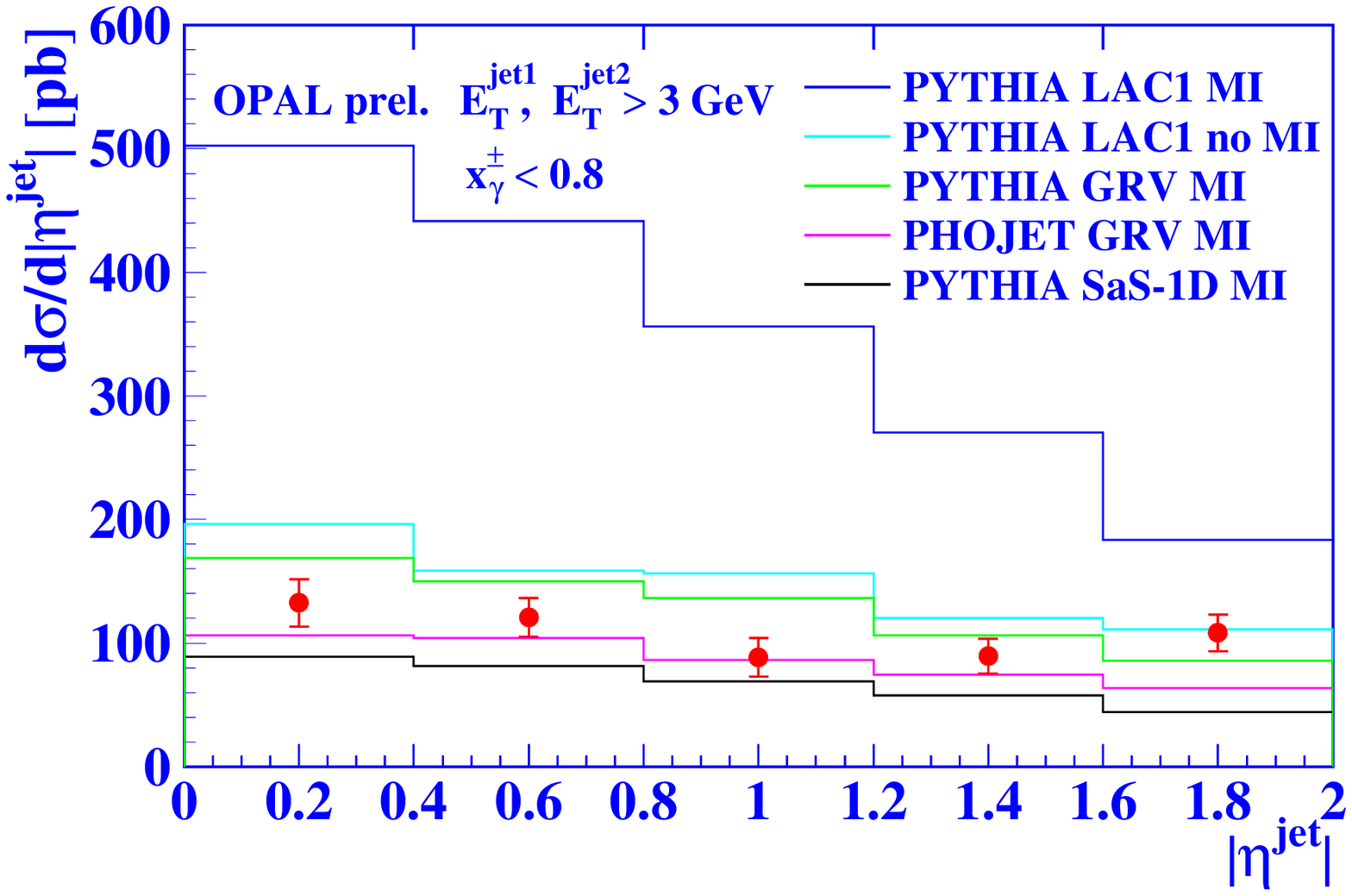,height=5.4cm,width=0.97\linewidth,clip=}}
  \caption{\label{fig:dijet} Dijet rate as a function of 
  pseudorapidity $\eta$~$^{\displaystyle {27)}}$.}

 \end{minipage}
\end{figure}

The rise of $F^{p}_2 (x,Q^2)$ at low $x$, as seen at HERA~\cite{Eisele}, 
has been shown to follow from normal DGLAP evolution~\cite{DGLAP}, with no 
need to invoke BFKL theory~\cite{BFKL}.  It is not obvious that the photon 
should behave in the same way, because of its  dual -- direct plus hadronic -- 
nature.  But the GRV parameterisation of the photonic parton densities does 
predict a very similar rise of $F^{\gamma}_2 (x,Q^2)$ at low $x$.  The LEP1 
results for $Q^2 \leq 6 GeV^2$ (Figure~\ref{fig:lowx}) 
are not in conflict with this.  
The more recent L3 result may appear to contradict OPAL~\cite{lowx}, 
but this is just 
another manifestation of the problem of unfolding $W_{\gamma \gamma}$, 
discussed above.  When L3 uses the same HERWIG Monte Carlo as used in the 
OPAL unfolding they get similar values of $F^{\gamma}_2 (x,Q^2)$.
The errors shown on both the OPAL and the L3 points are predominantly 
systematic and the difference between them merely reflects different choices 
from the range of available unfolding Monte Carlos.  Both experiments 
disagree with the much lower values of $F^{\gamma}_2 (x,Q^2)$ 
found in this region 
by the $TPC/2\gamma$ experiment~\cite{TPC}.  One should not, therefore, 
worry too much about the fact that the GRV curves are lower than 
the LEP points, since GRV was constrained by the $TPC/2\gamma$ data -- all 
that was then available.  

No new data can be expected from LEP2 in this low $Q^2$ region.  Both 
OPAL and L3 used their small-angle luminometers to tag the electrons in 
these events, and the definition of $Q^2$, above,  
tells us that at fixed $\theta_{tag}$ the mean value of 
$Q^2 \propto E_{beam}^2$.  So a definite statement about the low $x$ rise of 
$F^{\gamma}_2 (x,Q^2)$ is not waiting for new data.  It is waiting for 
the all-LEP working group  to sort out the 
$W_{\gamma \gamma}$ Monte Carlo modelling and 
unfolding problem so that we can reduce the systematic errors; 
very likely without much movement of the points already plotted in
Figures~\ref{fig:f2qcd} and ~\ref{fig:lowx}.

\subsection{QCD Gluons in the Photon }

Singly tagged $e \gamma$ scattering can never give direct measurements of the 
distributions of uncharged partons like gluons.  The only published attempts 
at a direct unfolding of $g^{\gamma}(x)$ come from H1~\cite{h1glue}, using 
the distributions of high-$p_T$ hadrons in photoproduction.  But the error 
bars are even larger than for the LEP measurements of 
$F^{\gamma}_2 (x,Q^2)$, and 
there is a fundamental problem with the method because the sample is 
contaminated by hadrons from ``underlying events'' in
which there are interactions of other 
partons from the colliding proton and target photon.  The uncertainty 
generated by such multiple interactions is the main reason why ZEUS has never 
attempted such an unfolding~\cite{JMBpc}.

Groups at both HERA~\cite{H1Z} and LEP are making progress in testing the 
parameterisations of $g^{\gamma}(x,Q^2)$ by studying the properties of high 
$E_T$ dijets as a function of variously defined ``$x_\gamma$'' variables.  A 
recent OPAL analysis of LEP2 data~\cite{Roland} uses an untagged 
$\gamma \gamma$ sample, with two identified high $E_T$ jets in which they 
define  $x_{\gamma}^+$ and $x_{\gamma}^-$ according to the formulae on 
Figure~\ref{fig:scat}.
For direct coupling of the two photons to two quarks
(Figure~\ref{fig:scat}a) we expect 
all of the hadrons to be in just two hard quark jets, so $x_{\gamma}^+$ 
and $x_{\gamma}^-$ should both be close to 1.
In a singly resolved event~\cite{Chrslls} 
there should be undetected hadronic momentum going into the forward region at 
one end of the detector, due to the unscattered remnant of the resolved 
photon  (Figure~\ref{fig:scat}b), so one or other of 
$x_{\gamma}^+$ or $x_{\gamma}^-$ 
should be significantly less than 1.  In doubly resolved events 
(Figure~\ref{fig:scat}c)) 
unscattered photon remnants will go into both forward regions, therefore both 
$x_{\gamma}^+$ and $x_{\gamma}^-$ should be much less than 1.  This picture 
is well confirmed, at least in the framework of the PYTHIA Monte Carlo 
program which allows events to be identified as coming from direct, singly 
resolved or doubly resolved processes (Figure~\ref{fig:scat}d).  
By making cuts on both 
of $x_{\gamma}^{\pm} \geq 0.8$ or one of $x_{\gamma}^{\pm} < 0.8$ in the real 
data, OPAL has separated a ``direct'' and a ``resolved'' 
sample.  Figure~\ref{fig:angular} shows 
the angular distributions in the dijet C. of M. for the two samples, compared 
with the predictions of various parton-parton scattering processes which 
match the ``resolved'' category, or of simple 
$\gamma \gamma \rightarrow q{\bar{q}}$ which matches the ``direct'' events.  
This work confirms the qualitative predictions of QCD, but the similarity of 
the different sub-processes in the resolved case is too close to give a 
measure of the gluon content.  A more quantitative test has been made by 
using various parton density functions with the PYTHIA and PHOJET Monte Carlo 
models to predict the differential jet-pseudorapidity distributions in high 
$E_T$ dijet events, Figure~\ref{fig:dijet}.  
But here again, as in the H1 attempt to 
unfold the gluon density, including the effects of multiple interactions 
(labelled ``MI'') makes a big difference, especially to the 
LAC1 case where the 
amount of glue in the photon is largest. \hfill

 \begin{wrapfigure}[16]{r}{8.4cm}
 \vspace{-0.1cm}
 \epsfig{file=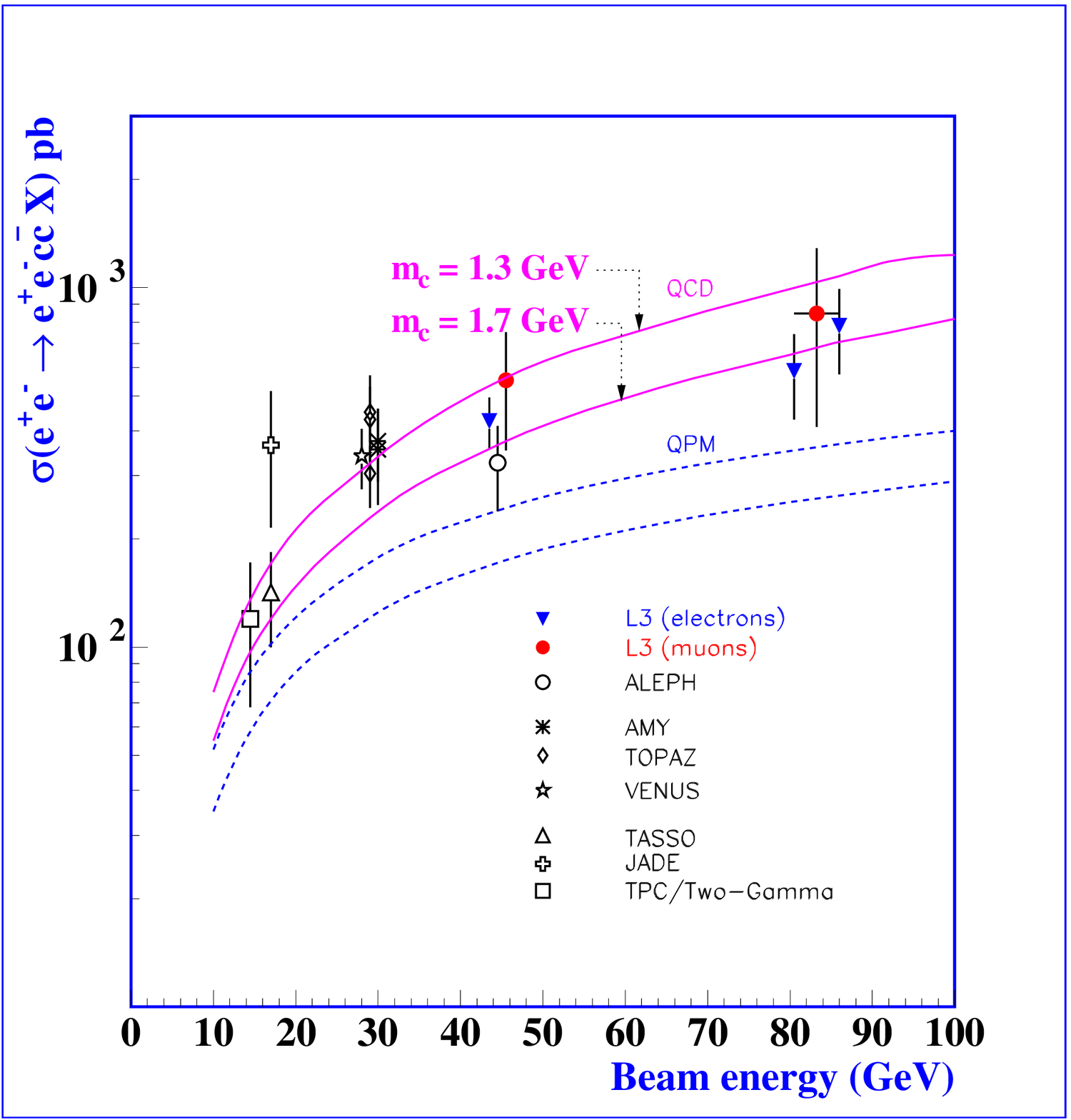,height=8.5cm,width=8.5cm,clip=}
 \vspace{-0.9cm}
 \caption{Cross section for $\gamma\gamma\rightarrow c\bar c$.}
 \label{fig:charm}
 \end{wrapfigure}

Two LEP2 untagged $\gamma \gamma$ analyses confirm the QCD picture, and may 
eventually allow a measurement of the glue.  
In the first, L3 has used high $p_T$ final state leptons to 
measure the inclusive charm production rate~\cite{L3D}, extending the 
work of the TRISTAN experiments and of ALEPH at LEP1 up to higher energies, 
Figure~\ref{fig:charm}.  The results can only be 
explained by including a large resolved 
contribution which grows with energy, as predicted by Kr\"amer et al. 
\cite{Kraemer+}  Deducing the precise amount of glue in the photon from 
this is difficult because, yet again, we do not know what value of $m_c$ 
to use -- hence the two lines for the ``QCD'' case which straddle most of the 
data points.  In the second, the $E_{T}^{jet}$ distributions from OPAL's 
dijet analysis have 
been compared with the predictions of a parton-level NLO calculation 
\cite{Kramer}.  Again, the resolved components are essential to explain the 
data.

\section{\bf Other QCD $\gamma \gamma$ Processes }
\subsection{Hadron $p_T$ Distributions}

\begin{wrapfigure}[20]{r}{8.6cm}
 \vspace{-1.4cm}
 \epsfig{file=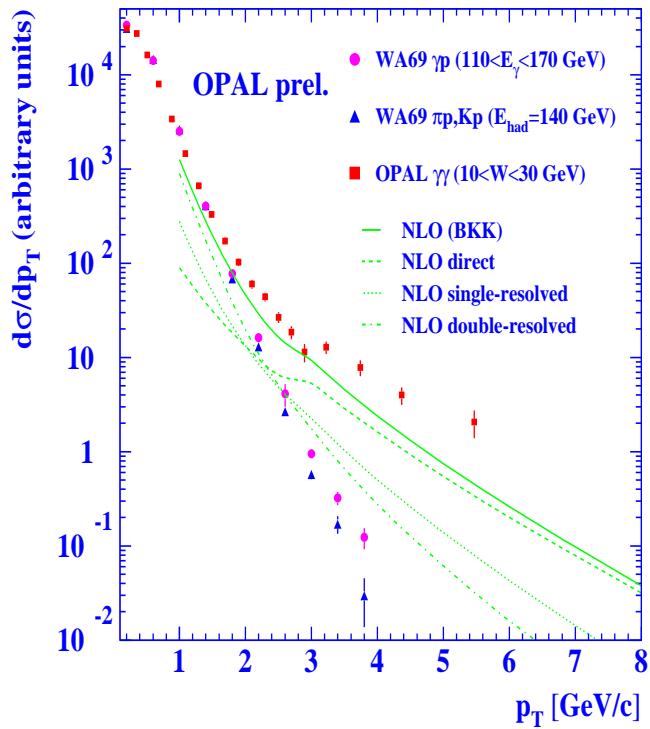,height=12.0cm,width=9.9cm,clip=}
 \vspace{-1.4cm}
 \caption{$p_T$ distribution of $hp$, 
$\gamma p$ and $\gamma\gamma$ scattering.}
 \label{fig:WA69}
 \end{wrapfigure}

The special nature of the photon shows up again in the OPAL~\cite{Stefan}
data of Figure~\ref{fig:WA69}, 
where the transverse momentum distributions of individual hadrons are shown 
for three different initial states; $\gamma \gamma$, $\gamma p$ and $\pi p$, 
all at the same C. of M. energy.  The distributions have been normalised to 
the same value at low $p_T$.  They look very similar in the soft region with 
$p_T < 1.4 GeV/c$, but the $\gamma \gamma$ data then diverge very markedly 
from the other two,  perhaps the clearest evidence yet for the direct 
$\gamma \gamma \rightarrow q \bar{q}$ process.  Checks from the other LEP 
experiments and comparisons from HERA are needed. 

\subsection{ Total Cross Section 
$\sigma ^{\gamma \gamma}_T (W_{\gamma \gamma})$ }

This has always been the most difficult number to measure in $\gamma \gamma$ 
physics.  Figures~\ref{fig:eflow}a) and b) 
above demonstrate for tagged events how the hadronic 
energy flow tends to go more and more forward, out of the efficient part of 
the detector, as the momentum transfer from the scattered electron and 
positron decreases.   But the biggest contribution to the total cross section 
is from untagged events where both parent leptons are forward scattered and 
go down the beampipe, undetected, with $P^2 \simeq Q^2 \simeq 0$.  And when 
both photons are close to the mass shell we expect a larger contribution from 
soft hadron-hadron scattering; diffractive processes with low $p_T$ final 
state hadrons which will mostly go into the badly measured regions.  This 
diffractive component must include the ``elastic scattering'' of the 
vector-meson-dominated photon, $\gamma \gamma \rightarrow \rho^0
\rho^0$, though 
no one has yet succeeded in detecting it at small angles at LEP. (The 
analogous $\gamma p \rightarrow \rho^0 p$ channel has been seen at HERA and 
behaves just as vector dominance and Regge factorisation would predict.)

Extraction of $\sigma ^{\gamma \gamma}_T (W_{\gamma \gamma})$ depends even 
more heavily upon unfolding than does the measurement of 
$F^{\gamma}_2 (x,Q^{2})$ described in section 2.1 above.  And the Monte 
Carlo programs that have to be used as input to the unfolding again differ 
seriously from one another.  OPAL sees particularly big differences between 
predictions from PHOJET and PYTHIA of the fraction of diffractive plus 
elastic events.

\begin{figure}[h]
 \vspace{-0.8cm}

 \hspace{-1.0cm}
 {\epsfig{file=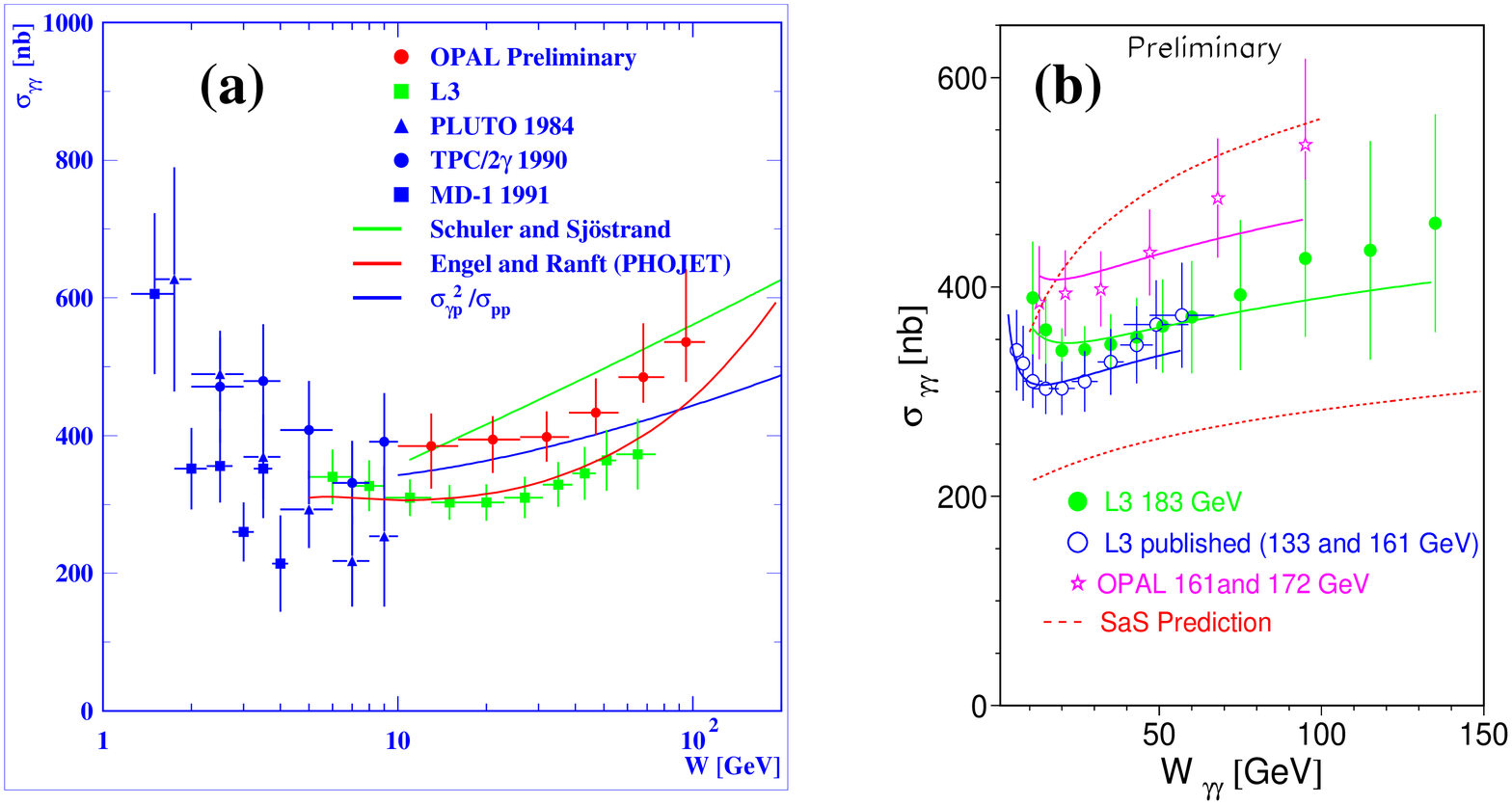,height=9.9cm,width=1.02\linewidth,clip=}}
  \caption{\label{fig:totx} Total $\gamma\gamma$ cross section .}

\end{figure}

Given the difficulties, it is not surprising that the first LEP measurements, 
from OPAL (preliminary~\cite{OPALsig}) and L3 (published~\cite{L3pub} and 
preliminary~\cite{Fredj}) disagree with one another, 
Figure~\ref{fig:totx}. Though the LEP 
measurements already agree better than did experiments at lower energy.  
Note that, because of unfolding, the LEP points at neighbouring values of 
$W_{\gamma \gamma}$ are very highly correlated.  Note also that the two L3 
results are significantly different from one another at low values of 
$W_{\gamma \gamma}$.  Systematic errors dominate, and are estimated by 
comparing the effects of unfolding with different Monte Carlo generators.  
OPAL reports that no generator has yet been found which correctly predicts 
the observed multiplicity distribution of hadrons, especially for low numbers 
of hadrons where the diffractive contribution should be important.  The 
message to theorists who want to use these results 
is that the spread of the data is a 
good indicator of the systematic errors, for the moment.  
Getting these errors down 
looks like a harder job for the all-LEP working group than sorting out 
$F^{\gamma}_2$.

One thing which both the OPAL and the L3 results have in common is a 
significant rise of $\sigma ^{\gamma \gamma}_T (W_{\gamma \gamma})$ 
between $W\simeq 20~GeV$ and $W \simeq 100~GeV$.  This rise is steeper than 
the simple factorisation prediction, labelled
$\sigma^2 _{\gamma p}/\sigma_{pp}$ on Figure~\ref{fig:totx}a), 
based on the published ZEUS and H1 
values for $\sigma_{\gamma p}$.  But a recent ZEUS thesis~\cite{Surrow} has 
extrapolated very low $Q^2$ photoproduction cross sections to $Q^2=0$.  The 
resulting estimates of the total cross section are significantly higher than 
the published values, and they rise faster with increasing $W_{\gamma p}$.  
This $\gamma p$ evidence, together with the OPAL and L3 data on 
$\sigma_{\gamma \gamma}(W_{\gamma \gamma})$, suggests that the total cross 
sections involving incoming photons are different from purely hadronic total 
cross sections.  In the familiar parameterisation~\cite{PDG} 
$\sigma_{tot}=Xs^{\epsilon}+Ys^{-\eta}$
Fredj~\cite{Fredj} has shown that the average of the OPAL and L3 results 
for $\sigma_{\gamma \gamma}(W_{\gamma \gamma})$ needs a larger exponent 
($\epsilon \simeq 0.16 \pm 0.02$) than the accepted value 
($\epsilon = 0.0790\pm 0.0011$) which fits all 
hadronic total cross sections.

\subsection{Doubly tagged $\gamma \gamma \rightarrow hadrons$. }

All of the LEP 
experiments are equipped with small calorimeters beyond the first mini-beta 
quadrupole, capable of tagging at at finite values, 
$P^2 \simeq 0.3 GeV^2$, of the virtuality of the target photon.
Rates at LEP1 for doubly tagged events were too low to be worthwhile 
(with only $\simeq 140 pb^{-1}$ of integrated luminosity), but preliminary 
tests at LEP2 are promising, and there are likely to be worthwhile  
measurements 
(with up to $500 pb^{-1}$) of the amount by which 
$F_{2}^{\gamma}(x,Q^2 , P^2 )$ is 
suppressed at finite $P^2$.  Others are searching 
for double tags at larger angles, looking for BFKL effects~\cite{deroecketc}.
No clear results have yet been reported.

\section{\bf $\gamma\gamma\rightarrow Resonances$ }

CLEO II has already collected  more than $3fb^{-1}$ integrated luminosity, and the 
beauty factory experiments BaBar and Belle expect even more.  Despite their 
disadvantage in energy compared with LEP they will always have access to 
larger samples of $\gamma\gamma\rightarrow resonances$, at least up to 
$m_{resonance} \simeq m_{\chi^{c}_2}$, and their final state hadrons are 
better measured because they are not as strongly boosted along the beam
direction as often happens at LEP.  So far CLEO II has published a few 
gems from its great treasure of data, and the LEP 
experiments have checked some of them.  

CLEO II's measurement of the partial width 
$\Gamma_{\eta_c \rightarrow \gamma \gamma} = 4.3 \pm 1.0 \pm 0.7 
\pm 1.4 $~\cite{Cleoetac} agrees 
with L3 and with the radically different Fermilab E760 measurement that uses 
$p \bar{p} \rightarrow \eta_c \rightarrow \gamma \gamma$.  But the E760  
value~\cite{fnal} for 
$\Gamma_{\chi_{c}^{2} \rightarrow \gamma \gamma} = 0.37 \pm 0.1 $ is 
significantly smaller than the values of around $1 keV$ measured by CLEO II  
and L3~\cite{ggchi}.  Both sides need to look again at their systematic 
errors.

The ``stickiness'' $S$ of a resonance X is defined as the phase-space weighted 
ratio of $\Gamma_{J/\psi \rightarrow X \gamma}$ to 
$\Gamma _{X \rightarrow \gamma \gamma}$.  Glueballs should be produced 
copiously in radiative $J/ \psi$ decays, but should not couple directly to 
$\gamma \gamma$.  For one particular glueball candidate, the $f_j (2220)$, 
CLEO II sees absolutely no signal, and has reported~\cite{Cleostick} a 
lower limit $S \geq 105$ -- the highest ever measured.  L3 also sees no sign 
of this resonance in $\gamma \gamma$, but with much lower 
significance~\cite{Wadhwa}.

CLEO II has also reported~\cite{Cleoff} singly tagged production of the 
three pseudoscalar mesons, $\gamma \gamma^* \rightarrow \pi^0$, $\eta$ or 
$\eta'$.  The $Q^2$ dependence is predicted to depend upon a dipole form 
factor, whose shape can be fitted to give an effective mass for the exchanged 
virtual vector meson.  CLEO obtains two roughly equal values of the
dipole mass, respectively 
$776\pm 10 \pm 12 \pm 6 MeV/c^2$ and $774 \pm 11 \pm 16 \pm 22$ for $\pi ^0$ 
and $\eta$ production, but a significantly larger value of 
$859 \pm 9 \pm 18 \pm 20 MeV/c^2$ for the $\eta'$ (L3 report $900 \pm 50$ 
for the $\eta'$~\cite{L3etapr}).  The size of the form 
factor at large $Q^2$ can be 
predicted from perturbative QCD models, assuming the $q \bar{q}$ wave 
function of the meson.  For $\pi^0$ and $\eta$ the value fits the simplest 
model already at $Q^2 =15 GeV^2$, but the $\eta'$ form factor is about twice 
the size predicted by the model.  Theorists~\cite{Brodsky} have suggested 
that this disagreement with pQCD, together with the higher dipole mass for 
the $\eta'$, mean that its wavefunction must be more complicated than those 
of the lighter psuedoscalars

\section{\bf Acknowledgements}

I am grateful for help and advice from colleagues in the all-LEP 
$\gamma \gamma$ working group, especially to Maria Kienzle and her L3
colleagues, and to my OPAL friends Richard Nisius and 
Stefan S\"oldner-Rembold. 
Jan Lauber at UCL has commented helpfully on my physics judgements and, 
with great patience, has put the figures into the text.


\begin{thebibliography}{99}
\bibitem{Kolan} H.~Kolanosky, Two Photon Physics at $e^+ e^-$ Storage 
Rings, Springer, Berlin, 1984.
\bibitem{Compton}
I.~Ginzburg, G.~Kotkin, V.~Serbo,V.~Telnov, Pizma ZhETF, {\bf 34}, 514 (1981)\\
I.~Ginzburg, G.~Kotkin, V.~Serbo,V.~Telnov, Nucl. Instr. and Meth., {\bf 205},
47 (1983)\\
V.~Telnov, Principles of Photon Colliders, in Proceedings of the Workshop on 
Gamma-Gamma colliders, LBL, Berkeley, March 28-31 1994; eds S.Chattopadhyay 
and A.M.Sessler, Nucl. Instr. and Meth. {\bf A355}, 1-194 (1995)\\
D.J.Miller, Other Options, in Proceedings of the Workshop on Physics and
Experiments with Linear Colliders, Morioka-Appi, Japan, 
September 1995, eds A.~Miyamoto, Y.~Fujii, T.~Matsui, S.~Iwata, 
World Scientific, Singapore, 322 (1996)
\bibitem{Borden} D.A.~Bauer, D.L.~Borden, D.J.~Miller, 
J.~Spencer, The Use of a 
Prototype Next Linear Collider for $\gamma \gamma$ and $e \gamma$ Collisions, 
 {\bf SLAC-PUB 5816}, June 1992. 
\bibitem{Miller86} D.J.~Miller, Can $F_L$ be measured? Proceedings of the 
ECFA Workshop on LEP 200, Aachen, September 1986, eds A.~B\"ohm, 
W.~Hoogland,  {\bf CERN 87-08}, 207-209 (1987)  
\bibitem{F2mu} 
R.~Akers{\em et al}, the OPAL collaboration, Z. Phys. 
{\bf C60}, 593 (1993)\\
P.~Abrue {\em et al}, the DELPHI collaboration, Z.Phys. {\bf C96}, 199 (1994)
\bibitem{L3mu} M.~Acciari {\em et al}, the L3 collaboration, 
{\bf CERN-EP/98-60}, sub. to Phys. Lett.{\bf B}
\bibitem{tau} R.~Akers {\em et al}, the OPAL collaboration, Z. Phys. 
{\bf C60}, 593 (1993)
\bibitem{L3QEDuntag}M. Acciarri {\em et al}, the L3 collaboration, 
Phys. Lett. {\bf B407},341 (1997)
\bibitem{FAFB} M. Acciarri {\em et al}, the L3 collaboration,
{\bf CERN-EP/98-06} sub. to  Phys. Letts.{\bf B}\\
C.A.~Brew for the ALEPH collaboration, in 
Proceedings of Photon '97,  10-15 May 1997, 
eds A.~Buijs, F.C.~Erne, World Scientific, Singapore (1997) 21.\\
K.~Ackerstaff {\em et al}, the OPAL collaboration, 
Zeits. Phys. {\bf C74}, 49  (1997) 
\bibitem{Witten}   E.~Witten, Nucl. Phys. {\bf B120}, 189 (1977) 
\bibitem{Fraser}   W.R.~Frazer, Phys. Lett.{\bf B194}, 287 (1987)
\bibitem{LEP200}   Physics at LEP 2, eds G.~Altarelli, 
T.~Sj\"ostrand, F.Zwirner, {\bf CERN 96-01} Vol I, 297-301.
\bibitem{F2LEP} 
K.~Ackerstaff {\em et al}, the OPAL collaboration, Z. Phys. 
{\bf C74}, 33 (1997)\\
K.~Ackerstaff {\em et al}, the OPAL collaboration,
Phys. Lett. {\bf B411} 387 (1997) 
P.Abreu {\em et al}, the DELPHI collaboration, Phys. Lett. {\bf B348},
 665 (1995)\\
A.~Finch, on behalf of the ALEPH collaboration, in Proceedings of Photon '97,
10-15 May 1997, eds A.~Buijs, F.C.~Erne, World Scientific, Singapore (1997)
\bibitem{BlobelKd}
V.~Blobel, Proceedings of the CERN School of Computing, Aiguablava, Spain, 
September 1984, {\bf CERN 85-09}, ed. C.~Verkerk.\\
A.~H\"ocker, V.~Karvelishvili, Nucl. Instr. Meth. {\bf A372} 469 (1996).\\
G.~D'Agostini, Nucl. Instr. Meth. {\bf A362} 487 (1995).
\bibitem{MC}{\bf ``PHOJET''} R.~Engel, Z.Phys. {\bf C66}, 203 (1995),\\ 
R.~Engel, J.~Ranft, Phys. Rev. {\bf D54}, 4246 (1996).\\
{\bf ``HERWIG''} G.~Marchesini, B.R.~Webber, Nucl. Phys. {\bf B238} 1 (1984)\\
G.~Marchesini {\em et al}, Comp. Phys. Comm. 
{\bf 67}, 465 (1992).\\
{\bf``PYTHIA''} T.~Sj\"ostrand, Comp. Phys. Comm {\bf 82}, 74 (1994) \\
{\bf``F2GEN''} A.~Buijs,W.J.G.~Langefeld, M.H.~Lehto, D.J.~Miller, Comp. Phys.
Comm. {\bf 79}, 523 (1994). 
\bibitem{MillerEg}D.J.~Miller, An Experimenter's Highlights, 
and J.A. Lauber {\em et al}, Tuning MC Models to fit DIS $e\gamma$ Scattering
Events, in Proceedings of Photon '97, 
10-15 May 1997, eds A.~Buijs, F.C.~Erne, World Scientific, Singapore
(1997) 52, 431
\bibitem{pdfs} M.~Gl\"uck, E.~Reya, A.~Vogt, {\bf ``GRV''}, Phys.Rev. 
{\bf D45} 3986 (1992) and Phys. Rev. {\bf D46}, 665 (1995).\\
H.~Abramowicz, K.~Carchula, A.Levy, {\bf ``LAC''}, Phys. Lett. {\bf B269}, 
458 (1991).\\
G.A.~Schuler, T.~Sj\"ostrand, {\bf ``SAS''}, Z. Phys. 
{\bf C68}, 607 (1995), and 
Phys. Lett. {\bf B376}, 193 (1996).
\bibitem{Millervogt}D.J.~Miller, A.Vogt, Kinematical Coverage for 
determining the Photon Structure function $F_2 ^{\gamma}$; in $e^+ e^-$
Collisions at TeV Energies, the Physics Potential, ed P.M.~Zerwas, 
{\bf DESY 96-123D} 473
\bibitem{Eisele} F.~Eisele, these proceedings.
\bibitem{DGLAP} {\bf ``DGLAP''} 
G.~Altarelli, G.~Parisi, Nucl. Phys. {\bf B126}, 298 (1977)\\
Yu.L.~Dokshitzer, Sov. phys. JETP, {\bf 46}, 641 (1977)\\
L.N.~Lipatov, Sov. J. Nucl. Phys. {\bf 20}, 95 (1975)\\
V.N.~Gribov, L.N.~Lipatov, Sov. J. Nuc. Phys. {\bf 15}, 438 (1972)  
\bibitem{BFKL} {\bf ``BFKL''}
E.A.~Kuraev, L.N.~Lipatov, V.S.~Fadin, Phys. Lett. {\bf 60B}, 50 (1975)\\
Ya.Ya.~Balitskij, L.N.~Lipatov, Sov. J. Nucl. Phys. {\bf 28}, 822 (1928)
\bibitem{lowx} K.~Ackerstaff {\em et al}, the OPAL collaboration, 
 Phys. Lett. {\bf B412}, 225 (1997)\\
M.~Acciarri {\em et al}, the L3 collaboration, {\bf CERN-EP/98-098}, 
sub. to Phys. Lett. {\bf B}.
\bibitem{TPC}  H.~Aihara {\em et al}, the $TPC/2\gamma$ collaboration,
Z. Phys. {\bf C34}, 1 (1987),\\
 Phys. Rev. {\bf D36}, 3506 (1987) 
\bibitem{h1glue} T.~Ahmed {\em et al}, the H1 collaboration, Nucl. Phys. 
{\bf B445}, 195  (1995)
\bibitem{JMBpc} J.M.~Butterworth, private communication.
\bibitem{H1Z}  J.~Breitweg {\em et al}, the ZEUS collaboration, 
Eur. Phys. J. {\bf C1}, 109 (1998)\\
C.~Adloff {\em et al}, the H1 collaboration, {\bf DESY 97-164}
\bibitem{Roland}K.~Ackerstaff {\em et al}, the OPAL collaboration, 
Zeits. Phys. {\bf C73}, 433 (1997)
and new result to be submitted to Eur. Phys. J. {\bf C}
\bibitem{Chrslls}C.H.~Llewellyn Smith, Phys. Lett. {\bf B79}, 83 (1978)
\bibitem{L3D} M.N.Kienzle-Focacci for the L3 collaboration, presented to
the conference DIS 98, Brussels 4-8 April 1998.
\bibitem{Kraemer+} M.~Drees, M.~Kr\"amer, J.~Zunft, P.M.~Zerwas, Phys. Lett. 
{\bf B306} 371 (1993), and Monte Carlo generator incorporating this 
model, written by the authors and A.~Finch.
\bibitem{Kramer} T.~Kleinwort, G.~Kramer, Phys. Letts. {\bf B370}, 141 (1996)
\bibitem{Stefan} S.~S\"oldner-Rembold for the OPAL collaboration, presented to 
the Heidelberg workshop 19-21 March 1998.
\bibitem{OPALsig} S.~S\"oldner-Rembold, Heidelberg workshop, {\em ibid} 
\bibitem{L3pub}   M.~Acciarri {\em et al}, the L3 collaboration,  
Phys. Lett. {\bf B408}, 450 (1997)
\bibitem{Fredj}   L.~Fredj for the L3 collaboration, presented to the
conference DIS 98, Brussels 4-8 April 1998.
\bibitem{Surrow} B.~Surrow, thesis, University of Hamburg, 
{\bf DESY-THESIS 1998-004}
\bibitem{PDG}Particle Data Group, Phys. Rev. {\bf D54}, 191-192 (1996)\\
A.~Donnachie,P.V.~Landshoff, Phys. Lett. {\bf B296}, 227 (1992) 
\bibitem{deroecketc} S.J.~Brodsky, F.~Hautmann, D.E.~Soper, in Proceedings
of Photon '97, 10-15 May 1997, eds A.~Buijs, F.C.~Erne, 
World Scientific, Singapore (1997) 100.\\
J.~Bartels, A.~De Roeck, C.~Ewerz, H.~Lotter, in ECFA/DESY
 Study on Physics and Detectors for a Linear Collider, ed. 
R.~Settles, {\bf DESY 97-123E}
\bibitem{Cleoetac} V.~Savinov, R.~Fulton,  for the 
CLEO II collaboration, in Proceedings
of Photon '95, 8-13 April 1995, eds D.J.~Miller, S.~Cartwright, V.~Khoze,  
World Scientific, Singapore (1995) 197.
\bibitem{fnal} T.A.~Armstrong {\em et al}, the E760 collaboration, Phys.
Rev. Letts. {\bf 70}, 2988 (1993)
\bibitem{ggchi} M.N.Kienzle-Focacci for the L3 collaboration, presented to
the conference DIS 98, Brussels 4-8 April 1998.
\bibitem{Cleostick} M.S.~Alam {\em et al}, the CLEO II collaboration, 
       Sub. to Phys. Rev. Lett. {\bf hep-ex/9805033}  
\bibitem{Wadhwa} M.~Wadhwa for the L3 collaboration, Recontres de
Moriond, 21-28 March 1998.
\bibitem{Cleoff}  J.~Gronberg {\em et al}, the CLEO II collaboration, 
Phys. Rev. {\bf D57}, 33 (1998)
\bibitem{L3etapr} M.~Acciarri {\em et al}, Phys. Lett. {\bf B 418}, 399 (1998)
\bibitem{Brodsky} S.J.~Brodsky, in Proceedings
of Photon '97, 10-15 May 1997, eds A.~Buijs, F.C.~Erne, World Scientific, 
Singapore (1997) 197;\\ 
S.~Ong, Photon '97 {\em ibid}, 272;\\ A.V.~Radyushkin, R.~Ruskov, 
  Photon '97 {\em ibid}, 277
\end{thebibliography}
\end{document}